\documentclass[lettersize,journal]{IEEEtran}
\usepackage{amsmath,amsfonts}
\usepackage{algorithmic}
\usepackage{algorithm}
\usepackage{array}
\usepackage[caption=false,font=normalsize,labelfont=sf,textfont=sf]{subfig}
\usepackage{textcomp}
\usepackage{stfloats}
\usepackage{url}
\usepackage{verbatim}
\usepackage{graphicx}
\usepackage{cite}
\usepackage{multirow,float,tipa, booktabs,amssymb,url,hyperref,tipa}
\usepackage[OT2,OT1,T1]{fontenc}
\usepackage{hyperref}  
\usepackage[usenames,dvipsnames]{color}
\hypersetup{colorlinks=true,pdfborder=0 0 0,citecolor=blue,linkcolor=blue,urlcolor=blue}

\newcommand{\RI}{\textcolor{black}}
\newcommand{\PDF}{\textcolor{black}}

\DeclareMathOperator*{\argmax}{arg\,max}

\begin{document}

\title{Pronunciation-Lexicon Free Training for Phoneme-based Crosslingual ASR via Joint Stochastic Approximation}

\author{Saierdaer Yusuyin, Te Ma, Hao Huang, ~\IEEEmembership{Member,~IEEE}, Zhijian Ou, ~\IEEEmembership{Senior Member,~IEEE}

\thanks{This work was supported in part by funding from TasiTech, in part by Guangxi Science and Technology Project under Grant (2022AC16002), and in part by the National Natural Science Foundation of China under Grant (62466055).
Corresponding author: Zhijian Ou, Hao Huang.}

\thanks{Saierdaer Yusuyin, Te Ma, Hao Huang are with the School of Computer Science and Technology, Xinjiang University, Urumqi 830046, China (e-mail: sar\_dar@foxmail.com; mate153125@gmail.com; huanghao@xju.edu.cn)}
\thanks{Zhijian Ou is with the Speech Processing and Machine Intelligence (SPMI) Lab, Department of Electronic Engineering, Tsinghua University, Beijing 100084, China, 
(e-mail: ozj@tsinghua.edu.cn)}
}



\maketitle

\begin{abstract}
Recently, pre-trained models with phonetic supervision have demonstrated their advantages for crosslingual speech recognition in data efficiency and information sharing across languages. However, a limitation is that a pronunciation lexicon is needed for such phoneme-based crosslingual speech recognition. 
In this study, we aim to eliminate the need for pronunciation lexicons and propose a latent variable model based method, with phonemes being treated as discrete latent variables. The new method consists of a speech-to-phoneme (S2P) model and a phoneme-to-grapheme (P2G) model, and a grapheme-to-phoneme (G2P) model is introduced as an auxiliary inference model. 
To jointly train the three models, we utilize the joint stochastic approximation (JSA) algorithm, which is a stochastic extension of the EM (expectation-maximization) algorithm and has demonstrated superior performance particularly in estimating discrete latent variable models.
\RI{Furthermore, we propose marginal likelihood scoring (MLS) decoding to align inference with the training objective and P2G  augmentation to improve the robustness of P2G mapping.}
Based on the Whistle multilingual pre-trained S2P model, crosslingual experiments are conducted in Polish (130 h) and Indonesian (20 h).
With only 10 minutes of phoneme supervision, the new method, JSA-SPG, achieves 5\% error rate reductions compared to the best crosslingual fine-tuning approach using subword or full phoneme supervision.
Furthermore, it is found that in language domain adaptation (i.e., utilizing cross-domain text-only data), JSA-SPG outperforms the standard practice of language model fusion via the auxiliary support of the G2P model by 9\% error rate reductions.
To facilitate reproducibility and encourage further exploration in this field, we open-source the JSA-SPG training code and complete pipeline.
\end{abstract}


\section{Introduction}
\label{sec:intro}
\IEEEPARstart{I}{n} recent years, deep neural networks (DNNs) based automatic speech recognition (ASR) systems, \PDF{which benefit from large amounts of transcribed speech data,} have made significant strides.
Remarkably, more than 7,000 languages are spoken worldwide \cite{Ethnologue}, and most of them are low-resourced languages.
A pressing challenge for the speech community is to develop ASR systems for new, unsupported languages rapidly and cost-effectively. Crosslingual ASR have been explored as a promising solution to bridge this gap \cite{Schultz98,XLSR,XLS-R,joinAP,liu2016adapting,liu2020multilingual}.

In crosslingual speech recognition, a pre-trained multilingual model is fine-tuned to recognize utterances from a new, target language, which is unseen \PDF{during} training the multilingual model. 
In this way, crosslingual speech recognition \PDF{can} achieve knowledge transfer from the pre-trained multilingual model to the target model, thereby reducing reliance on transcribed data and becoming \PDF{an} effective solution for low-resource speech recognition.
Most recent research on pre-training for crosslingual ASR can be classified into three categories - supervised pre-training with graphemic transcription or phonetic transcription, and self-supervised pre-training. The pros and cons of the three categories have recently been discussed in \cite{whistle}. 
Under a common experimental setup with respect to pre-training data size and neural architecture, it is further found in \cite{whistle} that when crosslingual fine-tuning data is more limited, phoneme-based supervised pre-training achieves the most competitive results and provides high data-efficiency.
This makes sense since phonetic units such as described in International Phonetic Alphabet (IPA), are \PDF{designed to represent speech} sounds shared in human language throughout the world. In contrast, the methods using grapheme units face challenges in learning shared crosslingual representations due to a lack of shared graphemes among different languages.

\PDF{We note an important distinction between phones (phonetic units) and phonemes (phonemic units). Producing accurate phonetic transcriptions requires substantial expert effort and is inherently subjective. Prior work—such as Whistle \cite{whistle} and ZIPA \cite{zhu2025zipa}—therefore relies on broad transcription, which operates at a coarser level of detail. While this coarsening may reduce phonetic precision, it enables stronger cross-lingual generalization. In our work, the goal is not to obtain highly precise phonetic labels. Instead, we use phonemes as an intermediate interface between speech and text to impose a useful structural constraint (an inductive bias), which we expect to reduce the overall complexity of the learning problem.}

A longstanding challenge in phoneme-based speech recognition is that phoneme labels are needed for each training utterance.
Phoneme labels are usually obtained by using a manually-crafted pronunciation lexicon (PROLEXs), which maps every word in the vocabulary into a phoneme sequence. Grapheme-to-phoneme (G2P) tools have been developed to aid this process of labeling sentences from their graphemic transcription into phonemes, but such tools are again created based on PROLEXs.
There are enduring efforts to compile PROLEXs and develop G2P tools \cite{novak2016phonetisaurus,mortensen2018epitran,hasegawa2020grapheme,deri2016grapheme,peters-etal-2017-massively,lee2020massively,li2022zero,zhu2022byt5} for different languages.
\PDF{Overall, existing phoneme-based ASR approaches rely heavily on manually curated PROLEXs or on G2P models trained from them, which limits their scalability—particularly when extending to truly low-resource languages.}

In this paper we are interested in reducing the reliance on PROLEXs in building phoneme-based crosslingual ASR systems, i.e., towards PROLEX free.
In recognizing speech $x$ into text $y$, \PDF{we assume that phonemes serve as intermediate states that structurally decompose the mapping.}
So we propose to treat phonemes as hidden variables $h$, and construct a latent variable model (LVM) with pairs of speech and text $(x,y)$ as observed values.
\PDF{The} whole model is a conditional generative model from Speech to Phonemes and then to Graphemes, which is referred to as a SPG model, denoted by $p_\theta(h,y|x)$.
SPG consists a speech-to-phoneme (S2P) model $p_\theta(h|x)$ and a phoneme-to-grapheme (P2G) model $p_\theta(y|h)$, and is thus a two-stage model\footnote{\PDF{Like in prior work on two-stage ASR as discussed in Section \ref{sec:related_work_two_stage}, the conditional independence $p_\theta(y|x,h)=p_\theta(y|h)$ is assumed, which is strictly not valid if considering prosodic effects of speech. This approximation currently does not pose a major issue for ASR performance, particularly in low-resource scenarios.}}.
Latent variable modeling enables us to train the SPG model, without the need to know $h$, by maximizing \PDF{the} marginal likelhood $p_\theta(y|x)$. This is different from previous two-stage ASR model\PDF{s} with phonemes as intermediate states, as reviewed later in Section \ref{sec:related_work}.
Learning latent-variable models usually involves introducing an auxiliary G2P model $q_\phi(h|y)$.

\textbf{Method contribution.}
Note that phonemes take discrete values, and recently the joint stochastic approximation (JSA) algorithm \cite{xu2016joint,ou2020joint} has emerged for learning discrete latent variable models with impressive performance.
In this paper, we propose applying JSA to learn the SPG model, which is called the JSA-SPG approach.
The S2P model is initialized from a pre-trained phoneme-based multilingual S2P backbone, called Whistle \cite{whistle}. 
Whistle generally stands for the \underline{w}eakly p\underline{h}onetic superv\underline{i}sion \underline{st}rategy for multilingua\underline{l} and crosslingual sp\underline{e}ech recognition\PDF{.} 
In supervised pre-training with phonemes, the phoneme labels are weak, \PDF{in that} they are obtained from G2P tools rather than human annotated.
Specifically, the Whistle model refers to the CTC-based S2P model pre-trained over 10 languages from CommonVoice \cite{ardila-etal-2020-common}, with a total of 4096 hours of training speech.
In applying JSA to SPG, when viewing phonemes as labels, we combine supervised learning over 10 minutes of transcribed speech with weak phoneme labels and unsupervised learning over a much larger dataset without phoneme labels.
That means we conduct semi-supervised learning. 
Bootstrapping from a good S2P backbone (like Whistle) and providing few-shots samples of latent variables (such as 10 minutes of weak phoneme labels) is found to be important to make JSA-SPG successfully work in the challenging task of crosslingual ASR.
\RI{Furthermore, we propose marginal likelihood scoring (MLS) decoding to align inference with the training objective and P2G  augmentation to improve the robustness of P2G mapping.
These components are important for making the overall JSA-SPG method work.}

\textbf{Experiment contribution.}
\PDF{Crosslingual experiments are conducted on Polish (130 h) and Indonesian (20 h) training data from the Common Voice dataset (v11.0) with grapheme transcriptions.}
Using only 10 minutes of phoneme supervision, JSA-SPG  outperforms the best crosslingual fine-tuning approach using subword supervision or full phoneme supervision.
\PDF{Furthermore, we investigate language domain adaptation, for which a standard approach is to train a new language model on the text data from the new domain. Unlike such standard approach, JSA-SPG can leverage the text-only data by using the G2P model to generate phoneme labels and adapt the P2G model, further improving cross-domain ASR performance.}
To promote future research along this direction, we release the JSA-SPG training code and complete pipeline at the following URL: \url{https://github.com/thu-spmi/CAT/tree/master/egs/JSA-SPG}

\section{Related Work}
\label{sec:related_work}
\subsection{Crosslingual ASR}
Multilingual and crosslingual speech recognition has been studied for a long time \cite{Schultz98}.
Modern crosslingual speech recognition typically fine-tunes a multilingual model pre-trained on multiple languages.
Most recent research on multilingual pre-training can be classified into three categories - supervised pre-training with graphemic transcription \cite{scalingMLASR,meta50,meta70,whisper,klejch22_Deciphering,liu2016adapting,liu2020multilingual} or phonetic transcription \cite{li2020universal,joinAP,tachbelie2022multilingual,saier}, and self-supervised pre-training \cite{XLSR,XLS-R,MMS}. 
It is shown in \cite{whistle} that when crosslingual fine-tuning data is more limited, phoneme-based supervised pre-training can achieve better results compared to subword-based supervised pre-training and self-supervised pre-training.
However, phoneme-based crosslingual fine-tuning in \cite{whistle} requires phoneme labels for every training utterance from the target language, which relies on a manually-crafted PROLEX for the target language.
The UniSpeech method \cite{wang2021unispeech} combines a phoneme-based supervised loss and a self-supervised contrastive loss to improve pre-training, and crosslingual fine-tuning still needs PROLEXs.
\RI{Recently, a line of research has explored zero-resource crosslingual ASR without any phone or grapheme transcribed supervision, notably the Deciphering Speech approach \cite{klejch22_Deciphering}. This work formulates ASR as a decipherment problem between unpaired speech and text, also adopting a cascaded S2P-P2G architecture. A universal phone recognizer trained on high-resource languages is employed for S2P, while the P2G conversion is modeled as an unsupervised decipherment using a noisy-channel formulation implemented with WFSTs. In this framework, the S2P model is fixed and only the P2G model is trained.
While both methods aim to remove pronunciation lexicons, JSA-SPG provides a more principled probabilistic formulation through marginal-likelihood training and decoding, jointly optimizing S2P and P2G. 
Another key distinction is that the predicted graphemes from the decipherment model \cite{klejch22_Deciphering} are used as pseudo-labels to train a separate grapheme-based acoustic model, whereas JSA-SPG directly enhances phoneme-based ASR through latent-variable learning.}

\subsection{Phoneme-based two-stage ASR}
\label{sec:related_work_two_stage}
The \PDF{two stages} of recognizing speech to phonemes and then to graphemes has been studied for crosslingual ASR \cite{xue2023tranusr,twostage}.
In \cite{twostage}, a two-stage crosslingual ASR architecture is used, which integrates an additional noise generator module during training to synthesize noisy phoneme sequences as P2G inputs, thus improving the robustness of the model.
Similarly, \cite{xue2023tranusr} implements a two-stage crosslingual ASR system, employing a pre-trained UniData2Vec model (a modified version of Unispeech\cite{wang2021unispeech}) as the phoneme recognition component coupled with a multilingual phoneme-to-word (P2W) converter.
The motivation of these works is similar to ours that phoneme units facilitate the learning of shared phonetic representations, making crosslingual transfer learning effective.
However, both studies require a PROLEX for the target language.

\subsection{Discrete latent variable models}
Hidden Markov models (HMMs) are classic discrete latent variable models (LVMs) and have been applied to ASR for a long time \cite{rabiner1989tutorial}.
Discrete LVMs are rarely used in recent end-to-end ASR systems, but have been widely used in many other machine learning applications such as dialog systems \cite{kim2020sequential,zhang-etal-2020-probabilistic}, program synthesis \cite{chen2021latent}, and discrete representation learning \cite{vqvae}.
A class of classic methods for learning LVMs consists of the expectation-maximization (EM) algorithm \cite{Dempster1977MaximumLF} and its extensions.
The joint stochastic approximation (JSA) algorithm \cite{xu2016joint,ou2020joint} is a stochastic extension of the EM algorithm with impressive performance.
Both the E-step and the M-step, which are often intractable in practice, are extended by the stochastic approximation methodology, hence called joint SA.
\PDF{Within the JSA framework, an inference model—serving as an adaptive proposal distribution for MIS sampling—is incorporated and continuously updated to provide diverse proposals for approximating the E-step. This adaptive procedure substantially improves sampling efficiency and is crucial for the strong empirical performance of JSA.}
JSA has been successfully applied to semi-supervised learning for semantic parsing \cite{song2020jae} and task-oriented dialog systems \cite{cai2022advancing,cai2023knowledge}.

\begin{figure*}[t]
  \centering 
  \includegraphics[width=0.8\textwidth]{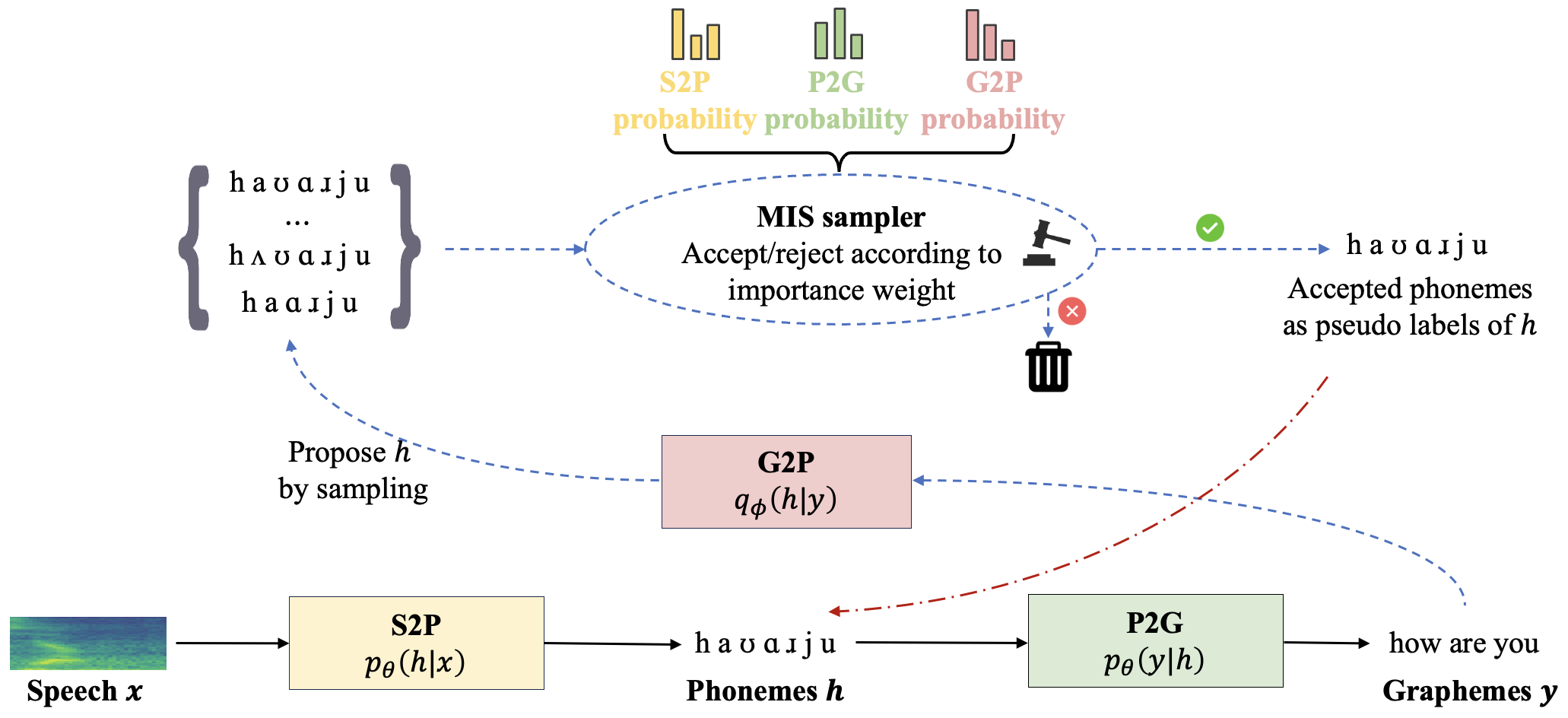}
  \caption{Overview of JSA-SPG.
1) The latent variable model (SPG) consists of speech-to-phoneme (S2P) and phoneme-to-grapheme (P2G). Learning SPG without knowing $h$ involves introducing an auxiliary G2P model.
2) During training, the G2P model proposes hypothesized phonemes, which get accepted or rejected according to the importance weight based on the probabilities calculated from the three model components.
The blue dashed line shows such Metropolis independence sampling (MIS), which is a Monte Carlo approximation of the E-step in EM.
3) The filtered phonemes are then treated as pseudo labels, as shown by the red dotted line.
4) Given the pseudo labels, we can calculate the gradients for the S2P model, the P2G model, and the G2P model, respectively, and proceed with parameter updating, very similar to perform supervised training, like the M-step in EM.}
  \label{fig:experiments}
\end{figure*}

\section{Method: JSA-SPG}
\label{sec:method}
\RI{In this section, we first formalize the SPG model and then describe the three core components of the JSA-SPG approach: JSA training, MLS (Marginal Likelihood Scoring) decoding, and P2G augmentation. We further present a case study on language domain adaptation to illustrate the additional benefits of the proposed JSA-SPG approach.}

\subsection{SPG model}
\label{ssec:model}
Let $(x,y)$ denote the pair of speech and text for an utterance. Specifically, $x$ represents the speech log-mel spectrogram and $y$ the graphemic transcription of $x$. 
Let $h$ denote the IPA phoneme sequence representing the pronunciation of $x$.
In recognizing speech $x$ into text $y$, we treat phonemes $h$ as hidden variables, and construct a latent variable model, which can be decomposed as follows:
\begin{displaymath}
    p_\theta(h,y|x) = p_\theta(h|x) p_\theta(y|h)
\end{displaymath}
Basically, as shown in Figure \ref{fig:experiments}, the whole model is a conditional generative model from Speech to Phonemes and then to Graphemes, which is referred to as a SPG model.
SPG consists a speech-to-phoneme (S2P) model $p_\theta(h|x)$ and a phoneme-to-grapheme (P2G) model $p_\theta(y|h)$.

\subsection{JSA Training}
\label{ssec:training}
Training the SPG model from complete data, i.e., knowing phoneme labels $h$, can be easily realized by supervised training.
To train S2P and P2G end-to-end (that is, conducting unsupervised training without knowing phoneme labels $h$), we resort to maximizing the marginal likelihood $p_\theta(y|x)$ and applying the JSA algorithm \cite{xu2016joint,ou2020joint}, which has emerged to learn discrete latent variable models with impressive performance.

The EM algorithm is a classic iterative method for finding maximum likelihood estimates of parameters in latent variable models. The EM algorithm iterates E-step (Expectation) and M-step (Maximization). The E-step needs to take expectation with respect to the latent variable, which is generally intractable for high-dimensional latent variable models. In this work, the latent variable is the phoneme sequence representing the pronunciation of a speech utterance, which is very high-dimensional. So, the EM algorithm can not be applied at all (not tractable) in this work.

JSA involves introducing an auxiliary inference model to approximate the intractable posterior $p_\theta(h|x,y)$, which, in the ASR task considered in this paper, is assumed to take the form of $q_\phi(h|y)$, i.e., a G2P model.
We can jointly train the three models (S2P, P2G and G2P), which is summarized in Algorithm \ref{alg:whistle-jsa} (JSA-SPG).
In our experiments, the S2P, P2G, and G2P models are all implemented—though not limited to—using the CTC model. In principle, any model compatible with Algorithm \ref{alg:whistle-jsa} can be employed in its place.

The JSA algorithm can be viewed as a stochastic extension of the well-known EM algorithm \cite{Dempster1977MaximumLF}, which iterates Markov Chain Monte Carlo (MCMC) sampling and parameter updating, being analogous to the E-step and the M-step in EM respectively.

\textbf{E-Step.}
The sampling step in JSA stochastically fills the latent variable $h$ (phonemes) through sampling from the posterior $p_\theta(h|x,y)$,
which is analogous to the E-step in EM (approximating the expectation via stochastic sampling).
However, direct sampling from the posterior $p_\theta(h|x,y)$ is intractable, so MCMC sampling is employed.
Particularly, using $p_\theta(h|x,y)$ as the target distribution and $q_\phi(h|y)$ as the proposal, we sample $h$ through Metropolis independence sampler (MIS) \cite{liu2001monte} as follows:

1) Propose $h' \sim q_\phi(h|y)$;

2) Let $h=h'$ with accept probability
$\min\left\lbrace 1, \frac{w(h)}{w(h^{\text{old}})} \right\rbrace$, and let $h=h^{\text{old}}$ otherwise,
where 
\begin{equation}\label{eq:impotance}
w(h) = \frac{p_\theta(h|x,y)}{q_\phi({h}|y)} \propto \frac{p_\theta(h|x)p_\theta(y|h)}{q_\phi({h}|y)}
\end{equation}
is the usual importance weight between the target and the proposal distribution and $h^{\text{old}}$ denotes the previous value for $h$ along the Markov chain.
In practice, we run MIS for several steps but for simplicity Algorithm \ref{alg:whistle-jsa} only shows a single step of MIS, \PDF{where} the chain is initialized from $p_\theta(h|x)$.

\textbf{M-Step.}
Once we obtain the sampled pseudo labels $\{ h^{(1)},h^{(2)},...,h^{(m)}\}$ from MIS, we can treat them as if being given and calculate the gradients for the S2P, P2G, and G2P models respectively and proceed with parameter updating, similar to the process in supervised training. 
This is analogous to the M-step in EM, but with the proposal $q_\phi$ being adapted as well.
In summary, the loss function can be written as:
\begin{equation}\label{JSA-RAG-loss}
\begin{aligned}
\mathcal{L}_{\text{JSA}} &=  -\frac{1}{m} \sum_{i=1}^m \left( \log p_{\theta}(h^{(i)}|x) \right. \\
& \left. + \log p_{\theta}(y|x,h^{(i)}) + \log q_\phi(h^{(i)}|y) \right)
\end{aligned}
\end{equation}

\begin{algorithm}[t]
    \caption{The JSA-SPG algorithm}
    \begin{algorithmic}
        \REQUIRE{S2P model $p_\theta(h|x)$, P2G model $p_\theta(y|h)$, G2P model $q_\phi(h|y)$, training dataset $\{(x,y)\}$}
        \REPEAT
        \STATE Draw a pair of speech and text $(x,y)$; 
            \STATE Initialize $h^{\text{old}}$ by sampling from $p_\theta(h|x)$;
            \STATE \textbf{\underline{Monte Carlo sampling:}}
            \STATE Sample $h'$ from the proposal $q_{\phi}(h|y)$;
            \STATE Let $h=h'$ (i.e., accept) with probability $ \min \left\{1,\frac{p_\theta(h|x)p_\theta(y|h)}{q_\phi(h|y)}/\frac{p_\theta(h^{\text{old}}|x)p_\theta(y|h^{\text{old}})}{q_\phi(h^{\text{old}}|y)} \right\}$, and let $h=h^{\text{old}}$ otherwise;
            \STATE \textbf{\underline{Parameter updating:}}
            \STATE Updating $\theta$ by ascending: ${\nabla}_\theta [p_\theta(h|x)  p_\theta(y|h)]$;
            \STATE Updating $\phi$ by ascending: ${\nabla}_\phi q_\phi(h|y)$;
        \UNTIL {convergence}
        \RETURN {$\theta$ and $\phi$}
    \end{algorithmic}
    \label{alg:whistle-jsa}
\end{algorithm}

\textbf{Semi-supervised Training.}
The JSA-SPG algorithm is general and is in fact an unsupervised learning over $(x,y)$ without the need to know $h$. It is challenging to run this purely unsupervised form from scratch in the ASR task considered in this paper, which involves very high-dimensional latent space. Two additional techniques are incorporated to add inductive bias into model training. 
\begin{itemize}
    \item First, the S2P model is initialized from a pre-trained phoneme-based multilingual S2P backbone, called Whistle \cite{whistle}, which has been shown to have good phoneme classification ability. In our experiment, JSA-SPG training with random initialization of the S2P model does not work well, especially when the amount of supervised phoneme data is small.
    \item Second, we assume that 10 minutes of transcribed speech with phoneme labels are available, which takes much less labor than compiling a complete PROLEX for a target language. Thus, we combine supervised learning over 10 minutes speech with phoneme labels and unsupervised learning over a much larger dataset without phoneme labels, i.e., conducting semi-supervised learning.
\end{itemize}
Bootstrapping from a good S2P backbone (Whistle) and providing few-shots samples of latent variables (10 minutes of phoneme labels) is found to be important to make JSA-SPG successfully work in the challenging task of crosslingual ASR. 

\textbf{Sampling from CTC based G2P and S2P.}
In JSA-SPG, we need to draw samples from the G2P model $q_{\phi}(h|y)$ as proposals for MIS. Sampling from the S2P model $p_{\theta}(h|x)$ is also required for initialization.
In this work, the G2P and S2P models are both implemented by Connectionist temporal classification (CTC) \cite{ctc}. 
For the CTC-based G2P model $q_{\phi}(h|y)$, assume that the length of $y$ is $T$.
At each position $t=1,\cdots,T$, we randomly draw a symbol (from the phoneme alphabet plus the blank symbol) based on the softmax distribution. In this way, we can draw a state path from the CTC model. Then, after reducing repetitive symbols to a single symbol and removing all blank symbols, we obtain a sample $h$ (i.e., a phoneme sequence) from $q_{\phi}(h|y)$. In practice, we independently draw multiple times at each position and obtain multiple independent state paths, which are converted to multiple $h$ samples.
Sampling from CTC-based S2P model $p_{\theta}(h|x)$ is similar.
Basically, the JSA-SPG algorithm can be applied to other types of models for G2P and S2P, such as attention-based encoder-decoder (AED) \cite{chorowski2014end} and RNN-transducer (RNN-T)  \cite{graves2012sequence}. 
We leave sampling from such models and apply the JSA-SPG algorithm for future work.

\textbf{More discussions on G2P and P2G models.}
\PDF{WFST-based methods have been developed for the G2P task \cite{novak2012improving,bisani2008joint}, known as joint-sequence models, which model the joint distribution $p(h, y)$ based on grapheme–phoneme joint multigrams.
Since these models are symmetric with respect to graphemes and phonemes in their WFST representations, a joint-sequence model can in theory also be applied to the P2G task.
This naturally raises the question of whether such joint-sequence models can be used within our SPG framework to realize both P2G and G2P.
However, it should be emphasized that the P2G task in SPG for speech recognition differs fundamentally from conventional P2G conversion tasks studied in \cite{bisani2008joint}.
In SPG, P2G aims to infer the grapheme sequence from the noisy hypothesized phoneme sequence produced by the S2P model, which may contain recognition errors, whereas traditional P2G focuses on mapping from clean, canonical phonemic transcriptions—making the SPG case considerably more challenging.}

\begin{figure*}
  \centering 
  \includegraphics[width=0.7\textwidth]{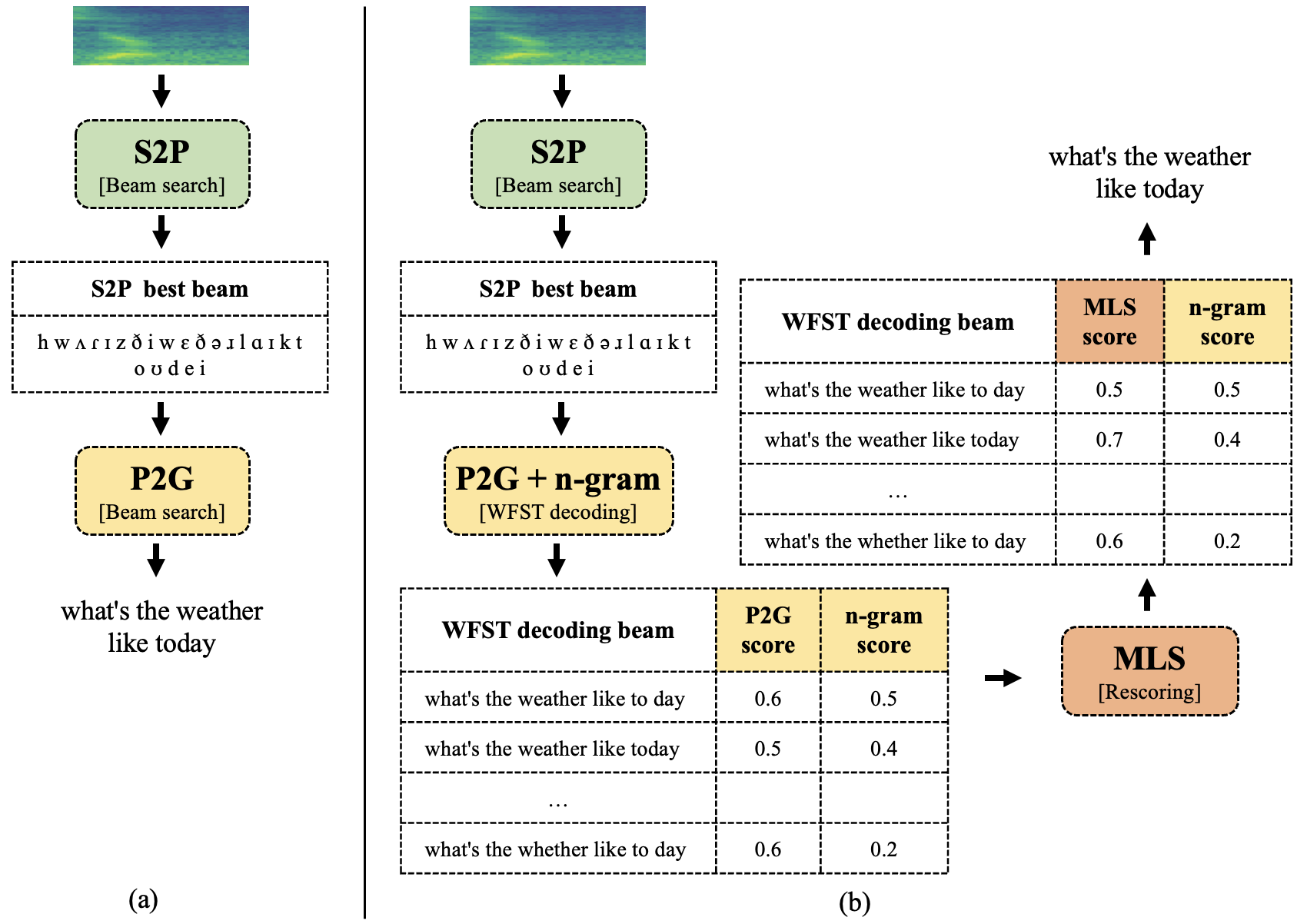}
    \caption{Illustration of decoding in JSA-SPG. (a) Vanilla decoding without LM; (b) Decoding with MLS rescoring. }
    \label{fig:decode}
\end{figure*}

\subsection{Decoding}
\label{ssec:decoding}
\textbf{Vanilla decoding.}
In testing, the S2P model first decodes out the phoneme sequence $h$ using beam search and selects the best beam $\hat{h}$ as input for the P2G model. Then, the P2G model also employs beam search to decode the graphemic result from $\hat{h}$, which is named as ``decoding without LM'' result. 
\PDF{In summary, the above steps can be written as:
\begin{equation}
\label{eq:vanilla}
\begin{aligned}
\hat{y} = &\argmax_y \sum_{h} p(y, h | x)
             = \argmax_y \sum_{h} p(y | h) p(h | x)\\
            \approx& \argmax_{y, h} p(y | h) p(h | x)
           \approx \argmax_{y} p(y| \hat{h})
\end{aligned}
\end{equation}
where $\hat{h} = \argmax_{h} p(h | x)$.}
Further, we train a word n-gram language model for WFST-based decoding from $\hat{h}$, which is named as ``decoding with LM'' result.
\PDF{This WFST-based decoding uses a subword lexicon, which maps from subwords to words, and does not require a pronunciation lexicon.}
Remarkably, the above two decoding procedures are vanilla approximations to the training objective (the marginal likelihood), which are collectively referred to as ``vanilla decoding''.

\textbf{Marginal likelihood scoring.}
\label{ssec:MLS}
Note that the training objective of the JSA algorithm is maximizing the marginal likelihood $p_\theta(y|x)$. 
So beyond of the vanilla decoding, we propose a new decoding algorithm, called ``decoding with marginal likelihood scoring'' (MLS). It consists of the following steps: 1) S2P takes in the audio $x$ and outputs the beam search best result $\hat{h}$; 2) P2G takes in the $\hat{h}$ and generates an n-best list of candidates $\hat{y}$ using WFST decoding; 3) G2P takes in each candidate hypothesis $\hat{y}$ and propose $k$ samples $\{ h^{(1)},h^{(2)},...,h^{(k)}\}$ from $q_\phi(h|\hat{y})$; 4) The marginal likelihood can be estimated with importance weights \cite{xu2016joint}, as shown in Eq. (\ref{eq:impotance}); 5) Each candidate hypothesis $\hat{y}$ is rescored using a sum of the estimated marginal likelihood and the weighted LM score. In summary, the above steps can be written as:
\begin{equation}\label{eq:decode}
\begin{aligned}
y^{\ast} = &\mathop{\arg\max}\limits_{\hat{y}}  \log \sum_{i=1}^{k}  \frac{p_\theta(h^{(i)}|x)p_\theta(\hat{y}|h^{(i)})}{q_\phi(h^{(i)}|\hat{y})} \\
&+ \lambda  \log p_\text{LM}(\hat{y})
\end{aligned}
\end{equation}
where $\hat{y}$ is taken from the n-best list from vanilla decoding, and $\lambda$ is LM weight.
There is a mismatch between vanilla decoding and the training objective. In contrast, MLS decoding maintains consistency with training by utilizing marginal likelihood scoring.
Additionally, note that vanilla decoding only uses the single best S2P result to fed to P2G for decoding, which is easily prone to error propagation. Decoding with MLS overcomes this drawback by scoring with multiple $h$.

\textbf{Improving P2G via data augmentation.}
\label{ssec:P2G}
Note that during the JSA-SPG training, as the models gradually converge, the diversity of phoneme sequences sampled by MIS decreases. The P2G model is gradually trained with less noisy input, compared with the input fed to P2G in testing.
In order to improve the robustness of the P2G model, we further augment the P2G model after the JSA-SPG training. Particularly, we decode 128 best phoneme sequences by S2P beam search decoding \RI{over the training speech data} and pair them with text labels, which serve as augmented data to further train the P2G model.

\subsection{Language Domain Adaptation}
\label{ssec:DomainAdaptation}
Note that after JSA-SPG training, we can use the auxiliary G2P model to generate phoneme labels on pure text. Below, we take the language domain adaptation task as an example to introduce the bonus brought by the G2P model.

Text-only data is easier to obtain than transcribed speech data. In cross-domain ASR, a common approach is to train external language models for language domain adaptation. 
In contrast, in JSA-SPG, we can use the G2P model to generate 64 best phoneme labels through beam search decoding, and then use the pairs of phonemes and text to continue adapting the P2G model. Then, we use the original S2P, the adapted P2G, and the cross-domain language model for speech recognition on cross-domain audio, which is found to outperform the standard practice of only doing language model fusion.


\section{Experiment}
\label{sec:experiment}
\subsection{Languages}
\label{ssec:languages}
In this paper, we aim to develop phoneme-based crosslingual speech recognition, in the scenario where a pronunciation lexicon is not available, by treating phonemes as discrete latent variables. Polish and Indonesian are two languages, which are outside of the ten languages in training the Whistle model and thus used for crosslingual speech recognition. Polish is close to the seen languages in Whistle model training, and there is a relatively large amount of training data (130 hours). On the other hand, Indonesian is not close to the seen languages and the data volume is small (20 hours). Therefore, our experiments on Polish and Indonesian represent two common scenarios in crosslingual speech recognition.

\begin{table*}
    \caption{\RI{Phoneme inventories of Polish and Indonesian. Uncovered phonemes denote those not included in the Whistle phoneme set.}}
    \label{tab:phone set}
    \begin{center}
    \begin{tabular}{l|l|l}
        \toprule
        Language & Phonemes & Uncovered phonemes\\
        \midrule
        Polish & \textscripta \ b \textctc \ d \texttoptiebar{dz} \texttoptiebar{d\textctz} \texttoptiebar{\textrtaild\textrtailz} \textepsilon \ f \textscriptg \ i \textbari \ j k l m n \textltailn \ \textipa{N} \textopeno \ p r s \textrtails \ t \texttoptiebar{t\textctc} \texttoptiebar{ts} \texttoptiebar{\textrtailt \textrtails} u v w x z \textrtailz \ \textctz & \texttoptiebar{d\textctz}  \texttoptiebar{\textrtaild\textrtailz} \texttoptiebar{t\textctc} \texttoptiebar{\textrtailt \textrtails}   \\
        Indonesian & \textscripta \ b d \texttoptiebar{d\textyogh} e \textschwa \ \textepsilon \ f \textscriptg \ \textramshorns \ h i \textsci \ j k l m n \textltailn \ \textipa{N} o \textopeno \ p q r s \textesh \ t \texttoptiebar{t\textesh} u \textupsilon \ v w x z  &    \\
        \bottomrule
    \end{tabular}
    \end{center}
\end{table*}

\subsection{Datasets}
\label{ssec:dataset}
\textbf{Common Voice}
\cite{ardila-etal-2020-common} is a large multilingual speech corpus, with spoken content taken primarily from Wikipedia articles. We conduct experiments on the Common Voice dataset released at September 2022 (v11.0). We select Polish (pl) and Indonesian (id) for JSA-SPG experiments, which were not used in Whistle pre-training. Polish has 130 hours \RI{(107K sentences)} of training data, while Indonesian has 20 hours \RI{(16.5K sentences)}, with an average sentence length of 4.3 and 4.5 seconds, respectively.  
In the JSA-SPG experiment, for each language of Polish and Indonesian, we used all its training data with the full set of text transcriptions, selected 100 utterances (about 10 minutes)  from the training set of each language and converted them into phonetic annotations using a publicly available phonemizer \cite{novak2016phonetisaurus}\footnote{The phonemizer developed in \cite{novak2016phonetisaurus} is called Phonetisaurus, which is a FST (Finite State Transducer) based G2P tool. The FSTs used in our experiments for Polish and Indonesian are from LanguageNet \cite{hasegawa2020grapheme}. The LanguageNet G2P models are available for 142 languages, with the phoneme error rates (PERs) ranging from 7\% to 45\%; but the particular PERs for Polish and Indonesian are not reported.}. 
\RI{
This is a simulated scenario for research purpose, where we use an available phonemizer.
It should be noted that except for generating the phoneme transcriptions for the selected 100 utterances, the phonemizer is no longer used in the remaining experimental steps.
In a practical scenario where neither a pronunciation lexicon nor a phonemizer is available, linguistic experts are required to manually annotate 100 utterances, which represents a moderate level of effort.
}

\RI{
Practically, linguistic experts can firstly identify the phoneme inventory for a target language (resources like Phobile\footnote{\RI{Phoible (\url{https://phoible.org/}) Release 2.0 from 2019 includes phoneme inventories for 2186 distinct languages.}} can be consulted for this purpose).
Then, linguistic experts can choose to annotate speech with good coverage of the phonemes appearing in the target language, especially for those phonemes that are not covered in the pre-training S2P backbone.
We establish two selection principles. First, ensuring that all phonemes in the target language are covered in the 10-minute annotated speech. Second, priority is given to select utterances that include those phonemes unseen by the pre-trained Whistle model.
The selection of the 100 utterances for both Polish and Indonesian in our experiments in this paper followed the same principles. Although the experiments were simulated, they can reasonably reflect the results expected in a realistic setup.
}

\textbf{VoxPopuli}
\cite{wang-etal-2021-voxpopuli} 
is a multilingual speech dataset of parliamentary speech in 23 European languages from the European Parliament. The Polish training set consists of 94.5 hours (or 0.7 M words) transcribed speech data, with an average sentence length of 10 seconds. We use the training set texts for language domain adaptation experiments. Additionally, the Polish validation set is used for model selection, and the test set is used for evaluation.

\textbf{Indonesian in-house data.}
We conducted Indonesian language domain adaptation experiments using our in-house dataset (VoxPopuli does not include Indonesian).
This dataset consists of 798 hours (or 6.16 M words) transcribed speech data, with an average sentence length of 5.18 seconds. We use the training set texts for language domain adaptation experiments. Additionally, the validation set is used for model selection, and the test set is used for evaluate the experimental results.

\subsection{Setup}
\label{ssec:setup}
\RI{For phoneme-based models, Table \ref{tab:phone set} lists the phoneme inventories for Polish and Indonesian, which were mainly derived from Phoible.
For each language, multiple phoneme inventories may exist, as documented in resources such as Phoible. By combining linguistic information from Wikipedia, subjective listening, and consultation with linguists, we finalize the phoneme inventory for each language, which is self-consistent with respect to our own inspection. The checking process for each language is detailed and archived at:
\url{https://github.com/thu-spmi/CAT/blob/master/egs/cv-lang10/lang-process/<lang>/lang_process.md},
where <lang> should be replaced with pl or id to access the corresponding documentation for Polish or Indonesian.}

\RI{The Polish inventory comprises 35 phonemes, with four absent from the Whistle phoneme set. The Indonesian inventory also includes 35 phonemes, all represented in Whistle.}
For subword-based models, both of the Polish and Indonesian alphabet size of subwords is 500. All text normalization and phonemization strategies are consistent with the Whistle work \cite{whistle}. For each language, we use its transcripts to separately train a word n-gram language model.
\PDF{For all results reported as ``decoding with LM'', WFST-based decoding is used.
WFST-based decoding for conventional phoneme-based systems (``Monolingual phoneme'', ``Whistle phoneme FT'', and ``Wav2vec2 phoneme FT'' in Table \ref{tab:jsa}), use a pronunciation lexicon, which maps from phonemes to words.
WFST-based decoding for SPG empolys a subword lexicon, which maps from subwords to words, and does not require a pronunciation lexicon.}

In the experiments, the S2P, P2G, and G2P models are all based on CTC. In CTC, the output sequence length needs to be smaller than the input length.
For the CTC based G2P model, the input is character\footnote{Polish uses 37 characters, and Indonesian uses 35 characters.} and the output is phonemes.
Note that the lengths of utterances in terms of characters are mostly longer than their corresponding phoneme sequences. For a very small portion of utterances with lengths that do not meet the CTC restrictions, we discard them.
For the CTC based P2G model, the input is phonemes and the output are BPEs. The BPE sequences of almost all data are shorter than their corresponding phoneme sequences, so they generally meet the length restrictions of CTC.

The Whistle-small 90M pre-trained model\footnote{\url{https://github.com/thu-spmi/CAT/tree/master/egs/cv-lang10/exp/Multilingual/Multi._phoneme_S}}
is used to initialize the S2P model. Both the G2P and P2G models use 8-layer Transformer encoders with dimension 512. We set the self-attention layer to have 4 heads with 512-dimension hidden states, and the feed-forward network (FFN) dimension to 1024. 
\RI{Both the G2P and P2G models have 18M parameters each.}
In the JSA-SPG training, for every data item, we obtain $m=10$ samples from G2P and run MIS for each sample. 
All experiment are taken with the CAT toolkit \cite{an2020cat}.
The learning rate for JSA-SPG is set to 3e-5, and when the validation loss does not decrease 10 epochs, the learning rate is multiplied by 0.5, training stop until it reaches 1e-6.
We extract 80-dimension FBank features from audio (resampled to 16KHz) as inputs to the S2P model. A beam size of 16 is used for S2P and P2G decoding in testing. $k=10$ samples from $q_\phi(h|y)$ is used for MLS decoding. We average the three best-performing checkpoints on the validation set for testing. 
\RI{During semi-supervised training, we train the model on the full set of training data with text transcriptions with a small subset of utterances with phoneme labels. To enhance exposure to phoneme labels, the phoneme-labeled utterances are oversampled by factors of 300 for Polish and 20 for Indonesian.}

\begin{table*}[ht]
    \caption{PERs (\%) and WERs (\%) for the experiments on \RI{Polish (130 h) and Indonesian (20 h)} from the Common Voice dataset. FT: fine-tuning. MLS: marginal likelihood scoring. $^\dagger$ denotes results from \cite{whistle}. NA denotes not applied. Except the column indicated by PER, other columns show WERs.}
    \label{tab:jsa}
    \begin{center}
    \begin{tabular}{l|c|c|cp{10mm}<{\centering}p{10mm}<{\centering}p{10mm}<{\centering}|cp{10mm}<{\centering}p{10mm}<{\centering}p{10mm}<{\centering}}
        \toprule
         \multirow{4}{*}{Exp.} & \multirow{4}{1.5cm}{\RI{Amount of \textit{text} ~~ transcription}} & \multirow{4}{1.5cm}{\RI{Amount of \textit{phoneme} transcription}} & \multicolumn{4}{c|}{Polish} & \multicolumn{4}{c}{Indonesian} \\
         & & & \multirow{3}{*}{PER}  & \multicolumn{3}{c|}{WER} & \multirow{3}{*}{PER} & \multicolumn{3}{c}{WER}  \\
         \cline{5-7}  \cline{9-11}
         & & &  & \multirow{2}{14mm}{decoding without LM} & \multirow{2}{11mm}{decoding with LM} & \multirow{2}{12mm}{decoding with \textit{MLS}} &   & \multirow{2}{14mm}{decoding without LM} & \multirow{2}{11mm}{decoding with LM} & \multirow{2}{12mm}{decoding with \textit{MLS}} \\
         & & && & & & & & &  \\
        \midrule
        Monolingual phoneme $^\dagger$ & \multirow{2}{*}{\RI{full}} & \RI{full} & 2.82 &  NA & 4.97 & NA & 5.74 & NA & 3.28 & NA \\
        Monolingual subword $^\dagger$ &  & \RI{NA} & NA &  19.38 & 7.12 & NA & NA  &  31.96 & 10.85 & NA \\
        \midrule 
        Whistle phoneme FT $^\dagger$ & \multirow{4}{*}{\RI{full}} & \RI{full} & \textbf{1.97} &  NA & 4.30 & NA & \textbf{4.79}  & NA & \textbf{2.43} & NA \\
        Whistle subword FT $^\dagger$ &  & \RI{NA} & NA & 5.84 & 3.82 & NA & NA &  12.48 & 2.92 & NA \\
        \RI{Wav2vec2 phoneme FT} $^\dagger$ &  & \RI{full} & 6.08 &  NA & 4.44 & NA & 6.30  & NA & 2.47 & NA \\
        \RI{Wav2vec2 subword FT} $^\dagger$ &  & \RI{NA} & NA & 7.49 & \textbf{3.45} & NA & NA &  12.01 & 3.15 & NA \\
        \midrule
        JSA-SPG & \multirow{2}{*}{\RI{full}} & \multirow{2}{*}{\RI{10 min}} & \multirow{2}{*}{17.35}  & \textbf{4.64} & 4.37 & \textbf{3.64} & \multirow{2}{*}{20.66}  & \textbf{4.55} & 2.92 & \textbf{2.31} \\
         ~~ w/o P2G augmentation &  &  &  & 8.19 & 4.65 & 3.93 &  & 9.04 & 3.26 & 2.47\\
        \bottomrule
    \end{tabular}
    \end{center}
    
\end{table*}

\section{Result and Analysis}
\label{sec:Result}
\subsection{Main Result}
\label{ssec:JSA-SPG-results}
\textbf{Monolingual baselines.}
For each language, there are two monolingual baselines, including monolingual phoneme-based training and subword-based training.
The phoneme-based training uses full phonetic annotations for 130 hours of Polish and 20 hours of Indonesian data.
The results are taken from \cite{whistle} and shown in the first and second rows of Table \ref{tab:jsa}.

\textbf{\RI{Whistle based} crosslingual fine-tuning baselines.}
There are two end-to-end crosslingual baselines for each language, both based on Whistle, the pre-trained phoneme-based multilingual S2P backbone.
The third and fourth rows in Table \ref{tab:jsa} show their results, respectively, which are also taken from \cite{whistle}.
Specifically, the phoneme-based Whistle-small pre-trained model is end-to-end fine-tuned with phoneme labels or subword labels for crosslingual speech recognition, which correspond to ``Whistle phoneme FT'' and ``Whistle subword FT'' in Table \ref{tab:jsa} and represent the two state-of-the-art crosslingual fine-tuning approaches.
Phoneme fine-tuning used full phonetic annotations.
Remarkably, in Indonesian, ``Whistle phoneme FT'' outperforms ``Whistle subword FT'', whereas in Polish, the opposite is observed.
As analyzed in \cite{whistle}, when crosslingual fine-tuning data are more limited (Indonesian has 20 hours of data vs Polish 130 hours), phoneme-based fine-tuning is more data-efficient and performs better than subword fine-tuning.

\RI{\textbf{Wav2vec2 based crosslingual fine-tuning baselines.}
Cross-lingual fine-tuning results of Wav2Vec 2.0 for each language are listed in the fifth and sixth rows of Table \ref{tab:jsa}, which are also taken from from \cite{whistle}.
The Wav2vec 2.0 model \cite{baevski2020wav2vec} are pre-trained on the same multilingual dataset of ten languages as the Whistle model. Note that the pre-trained model serves solely as an acoustic encoder. To enable crosslingual speech recognition, we introduce a linear layer followed by softmax on top of the encoder output and perform full-parameter fine-tuning using labeled data from the target languages (Polish and Indonesian). The fine-tuning labels are provided in the form of either phonemes or subwords, corresponding to the configurations ``Wav2vec2 phoneme FT'' and ``Wav2vec2 subword FT'' in Table \ref{tab:jsa}.
}

\textbf{JSA-SPG crosslingual ASR experiment with 10-minute phoneme labels.}
In the following, we introduce the JSA-SPG crosslingual ASR experiments with only 10 minutes of data per language having phoneme annotations.

In JSA-SPG, we first fine-tune the Whistle model on 10 minutes of phoneme labels to initialize the S2P model. Subsequently, this S2P model is utilized to generate phoneme pseudo-labels on the full training set, which are then used to train the P2G and G2P models for initialization. Ater such a initialization, the JSA-SPG algorithm is employed to jointly train the three models (S2P, P2G and G2P). \RI{``JSA-SPG'' refers to our overall method, which integrates the three core components: JSA training, MLS decoding, and P2G augmentation. 
More analysis results are presented in Section \ref{ssec:analysis}.}

For the results in Table \ref{tab:jsa}, JSA-SPG consistently achieves 5\% error rate
reductions compared to the best crosslingual
fine-tuning approach using subword or full
phoneme supervision (3.82 vs 3.64 for Polish, 2.43 vs 2.31 for Indonesian). Some detailed observations are as follows:

\emph{1) Compared to ``Whistle phoneme FT'',} the training of the three models (S2P, P2G and G2P) in JSA-SPG is trained in a more end-to-end way by maximizing marginal likelihoods over pairs of speech and text, hence obtaining better performance in lower WERs.
Full supervision of weak phoneme labels produces worse results, presumably because the full phoneme annotations may contain errors.
``Whistle subword FT'' is trained in an end-to-end way. 

The PERs of the S2P model in JSA-P2G are higher compared to the fully supervised approach ``Whistle phoneme FT''.
Remarkably, this work does not aim to obtain accurate phonetic transcription. Additionally, it is worthwhile to highlight that the PERs are significantly reduced by using SPG-JSA with only 10 minutes of phoneme labels. Originally, the Whistle model obtains PERs of 58\% and 46\% for Polish and Indonesian respectively. After using SPG-JSA with only 10 minutes of phoneme labels, the PERs are  dramatically reduced to 17.35\% and 20.66\% for Polish and Indonesian respectively.

\emph{2) The inferior results of ``Whistle subword FT'' compared to JSA-SPG} indicates the advantage of the SPG architecture (the benefit of explicitly modeling phonemes). The fine-tuning of Whistle using subword labels breaks the structure imposed by SPG.
Introducing phonemes as an intermediate interface between speech and language in the SPG architecture provides a useful structural constraint (inductive bias), which hopefully can reduce the complexity of the overall learning problem.
\PDF{A potential concern} is that the SPG pipeline contains discretized phonemes and gradients cannot propagate from P2G to S2P. However, the optimization of JSA-SPG is actually in an end-to-end way -- JSA is powerful in learning discrete latent variable models, propagating gradients across the P2G and S2P components, which is different from optimizing a single neural network but turns out to be very effective.

\RI{
\emph{3) The Wav2Vec2 based fine-tuning results} exhibit a trend similar to the Whistle based fine-tuning results, in comparing ``phoneme FT'' and ``subword FT''. For Polish, which has more training data (130h), the Wav2Vec2 based subword fine-tuning approach performs better than phoneme fine-tuning, and even slightly outperforms the JSA-SPG model. Nevertheless, in the low-resource scenario such as for Indonesian (20h), phoneme-based fine-tuning demonstrates a clear advantage over subword fine-tuning, and the JSA-SPG model significantly surpasses the Wav2Vec2 based models.
}

\RI{
\emph{4) MLS decoding and P2G augmentation are integral components of the JSA-SPG approach, playing important roles in its performance gains.}
For Polish, when trained without P2G augmentation and using vanilla decoding, the raw JSA-SPG model obtains a WER of 4.65\%. Incorporating P2G augmentation reduces to the WER to 4.37\%, and applying MLS decoding further improves it to 3.64\%.
The results are similar for Indonesian.
Vanilla decoding, as formulated in Eq. (\ref{eq:vanilla}), is suboptimal for the two-stage JSA-SPG model. These results demonstrate that the proposed MLS decoding effectively aligns the inference process with the training objective, while P2G augmentation enhances the robustness of the P2G mapping.
}
\begin{table*}
    \caption{WERs (\%) of cross-domain language domain adaptation (LDA) experiments from CV to VoxPopuli Polish and in-house Indonesian datasets. The FT denotes fine-tuning. The MLS denotes marginal likelihood scoring.}
    \label{tab:adaptation}
    \begin{center}
    \begin{tabular}{l|ccc|ccc}
        \toprule
        \multirow{3}{*}{Exp.} & \multicolumn{3}{c|}{Polish} & \multicolumn{3}{c}{Indonesian} \\
         & \multirow{2}{14mm}{decoding without LM} & \multirow{2}{11mm}{decoding with LM} & \multirow{2}{12mm}{decoding with \textit{MLS}} & \multirow{2}{14mm}{decoding without LM} & \multirow{2}{11mm}{decoding with LM} & \multirow{2}{12mm}{decoding with \textit{MLS}} \\
         & & & & & &  \\
        \midrule
        Whistle subword FT on CV & 33.46 & 22.58 & NA & 43.69 & 12.39 & NA\\
        \midrule
        JSA-SPG on CV & 35.18 & 29.04 & 26.79 & 39.19 & 16.93 & 14.28 \\
        ~~~~ + LDA training & 28.87 & 23.84 & \textbf{20.57} & 30.69 & 12.68 & \textbf{11.23} \\
        \bottomrule
    \end{tabular}
    \end{center}
\end{table*}

\begin{figure*}
\centering
  \includegraphics[width=0.5\textwidth]{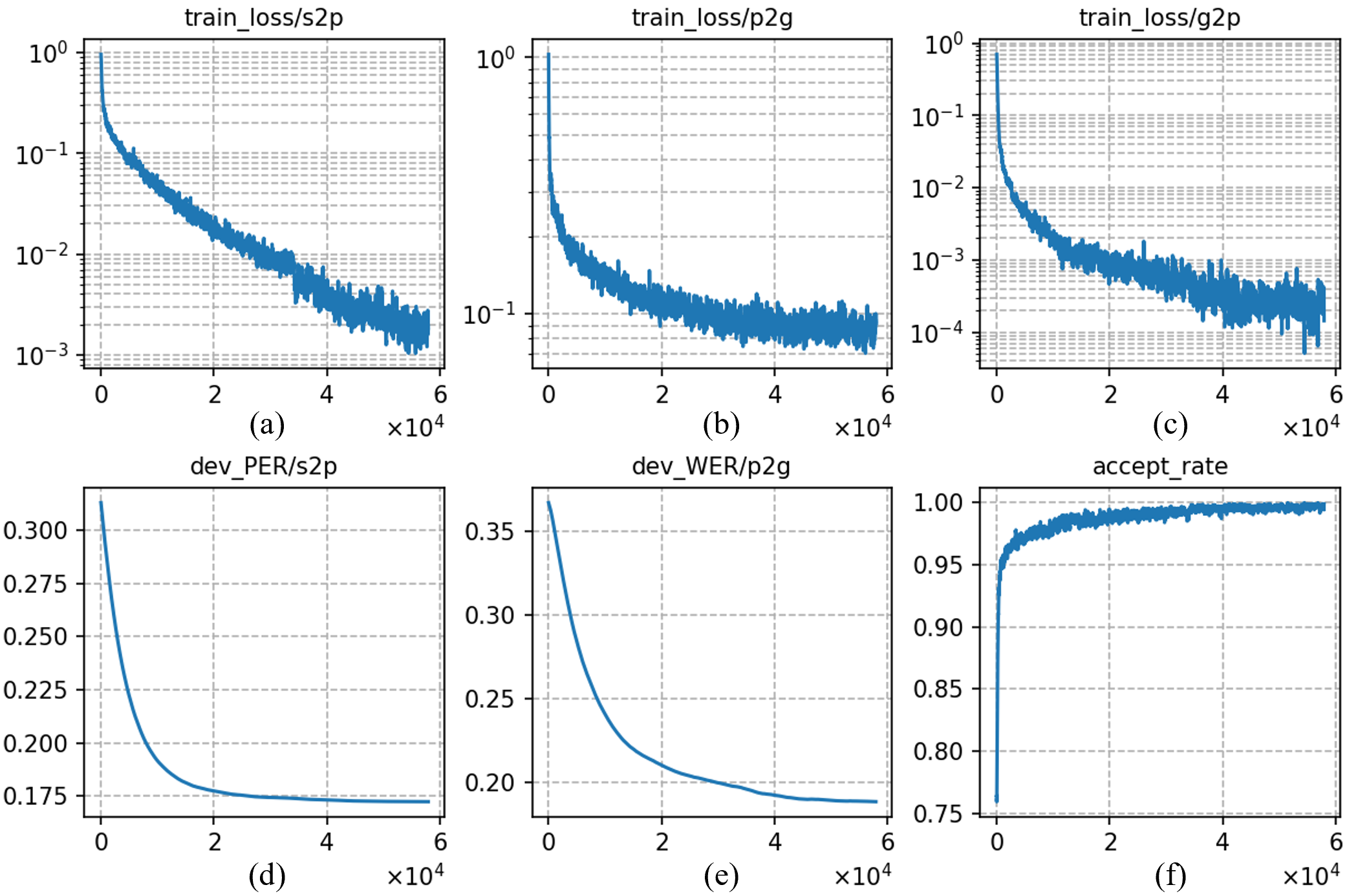}
    \caption{Plots of training and validation curves in JSA-SPG training on Common Voice Polish data. (a), (b), (c) represent the train losses of the S2P, P2G, and G2P models in the JSA-SPG training, respectively. (d) and (e) are the error rates of S2P and P2G models in the validation set. (f) represents the ratio of the number of samples accepted by the MIS sampler to the total number of samples proposed by G2P in one iteration.}
\label{fig:curve}
\end{figure*}

\subsection{Language Domain Adaptation Result}
\label{ssec:out-domain-results}
As shown in Table \ref{tab:adaptation}, for Polish, we test our models on VoxPopuli Polish test set, while both Whistle and JSA-SPG models is train on the Common Voice dataset. The CommonVoice dataset is comprised of texts from Wikipedia, recorded by users on mobile devices, while the VoxPopuli dataset consists of audio recordings of speeches from the European Parliament. Notably, 61.5\% of the words in the VoxPopuli Polish training set do not appear in the CommonVoice vocabulary list, and 31.5\% of the words in the test set are also absent. This indicates significant differences between the two datasets in terms of linguistic context, vocabulary, recording equipment, and average sentence length. We only use the text from the VoxPopuli training set and train a word-level 4-gram language model for language model fusion. 

The first row in Table \ref{tab:adaptation} shows the results of testing the Whistle subword fine-tuning model directly with cross-domain language model integration, which is a common method used in cross-domain speech recognition. Next, we test the JSA-SPG model directly without further training (the second row of Table \ref{tab:adaptation}). We then apply the domain adaptation method, introduced in Section \ref{ssec:DomainAdaptation}, to continue training the P2G model on VoxPopuli training text, and the result is shown in the third row.
It can be seen that after LDA training of P2G, the JSA-SPG model achieves significant improvements on cross-domain ASR tasks, indicating its strong extendibility.
Compared to the standard practice of language model fusion, we can also clearly see the advantage of JSA-SPG with LDA training, and its performance far exceeds that of traditional language domain adaptation method by 9\% error rate reduction (22.58 vs 20.57).

For Indonesian, our in-house Indonesian dataset is from audio books, which has a clear domain difference from the CommonVoice dataset.
Indonesian experiments are taken simlarly to Polish.
The JSA-SPG model with the adapted P2G model obtains the best result, significantly outperform Whistle subword fine-tuning model by 9\% error rate reduction (12.39 vs 11.23) as well.

\begin{table*}[ht]
    \caption{\RI{PER (\%) and WER (\%) for comparing JSA-SPG to SPG (Speech-to-Phoneme-to-Grapheme) based ASR baselines. MLS: marginal likelihood scoring.}}
    \label{tab:ablation}
    \begin{center}
    \begin{tabular}{p{40mm}|p{13mm}<{\centering}|p{4mm}<{\centering}p{12mm}<{\centering}p{10mm}<{\centering}p{10mm}<{\centering}|p{4mm}<{\centering}p{12mm}<{\centering}p{10mm}<{\centering}p{10mm}<{\centering}}
        \toprule
          \multirow{4}{*}{Exp.} & \multirow{4}{1.5cm}{Amount of \textit{phoneme} transcription} & \multicolumn{4}{c|}{Polish} & \multicolumn{4}{c}{Indonesian} \\
          & & \multirow{3}{*}{PER}  & \multicolumn{3}{c|}{WER} & \multirow{3}{*}{PER} & \multicolumn{3}{c}{WER}  \\
         \cline{4-6}  \cline{8-10}
          & &  & \multirow{2}{14mm}{decoding without LM} & \multirow{2}{11mm}{decoding with LM} & \multirow{2}{12mm}{decoding with \textit{MLS}} &   & \multirow{2}{14mm}{decoding without LM} & \multirow{2}{11mm}{decoding with LM} & \multirow{2}{12mm}{decoding with \textit{MLS}} \\
          & & & & & & & & &  \\
        \midrule
        SPG seperate training & \multirow{2}{*}{full} & \multirow{2}{*}{\textbf{1.97}}  & 10.16 & 4.77 & 4.64 &  \multirow{2}{*}{\textbf{4.79}}  & 14.28 & 4.34 & 2.99 \\
         ~~ + P2G augmentation  &  &   & 5.67 & 4.86 & 4.63 &  & 7.29 & 3.47 & 2.69 \\
        \midrule
        SPG training init from G2P & \multirow{2}{*}{10 min} & \multirow{2}{*}{17.72}  & 8.73 & 4.68 & 5.91 &  \multirow{2}{*}{21.85}  & 10.15 & 3.81 & 3.09 \\
         ~~ + P2G augmentation &  &   & 5.93 & 4.97 & 5.88 &  & 6.34 & 3.44 & 2.91 \\
        \midrule
         JSA-SPG w/o P2G augmentation & \multirow{2}{*}{10 min} & \multirow{2}{*}{17.35} & 8.19 & 4.65 & 3.93 & \multirow{2}{*}{20.66} & 9.04 & 3.26 & 2.47\\
         ~~ + P2G augmentation & &   & \textbf{4.64} & \textbf{4.37} & \textbf{3.64} &   & \textbf{4.55} & \textbf{2.92} & \textbf{2.31} \\
        \bottomrule
    \end{tabular}
    \end{center}
    
\end{table*}

\subsection{Analysis}
\label{ssec:analysis}
\textbf{Training process of JSA-SPG.}
To provide an intuitive understanding of the JSA-SPG training process, Figure \ref{fig:curve} shows the changes in several key indicators over the number of training iterations. It can be seen that the training losses of all three models and the validation error rates gradually decrease when using the JSA-SPG Algorithm \ref{alg:whistle-jsa}, clearly showing the ability of JSA-SPG for model optimization. 
Through JSA-SPG training, compared to the model fine-tuned with only 10 minutes of phonetic labels, which is the initial model in the experiment, the JSA-SPG model achieves a relative PER reduction of 45\% and a WER reduction of 48\% on the validation set.

\RI{
\textbf{Comparing JSA-SPG with other SPG based ASR baselines that incorporate P2G as an auxiliary task.}
We examine two SPG based ASR baselines that also incorporate P2G as an auxiliary task.
In the first baseline, referred to as ``SPG separate training'', the S2P, P2G and G2P components are separately trained using the full set of phoneme labels.
The second baseline, similar to JSA-SPG, uses only 10 minutes of phoneme labels together with the full set of text labels, but employs a different initialization scheme.
The results are summarized in Table \ref{tab:ablation}, where the last two rows are reproduced from the last two rows of Table \ref{tab:jsa} for ease of reference.}

\RI{
The first is a seemingly strong baseline in the context of training SPG models.
It represents a common practice for training SPG models, where a fixed, pretrained multilingual G2P model is used to generate phoneme labels for the target languages\footnote{\RI{A number of G2P tools are available for this purpose \cite{novak2016phonetisaurus,mortensen2018epitran,hasegawa2020grapheme,deri2016grapheme,peters-etal-2017-massively,lee2020massively,li2022zero,zhu2022byt5}. In our experiment, the Phonetisaurus tool with LanguageNet FSTs is used to generate phoneme labels \cite{novak2016phonetisaurus,hasegawa2020grapheme}.
It is interesting to examine the effects of different G2P tools, which is outside the scope of this paper.}}.
From Table \ref{tab:ablation}, we can see that due to errors in phoneme labels, using the full set of phoneme labels performs worse than JSA-SPG with 10 minutes of phoneme annotations. This demonstrates the power of the end-to-end optimization enabled by the JSA algorithm, which can take advantage of both the SPG structure and the training data very well with limited (10 minutes) phoneme annotations. Remarkably, for ``SPG separate training'', the improvement brought by MLS decoding over vanilla decoding is not as large as for JSA-SPG. Presumably, this is because MLS decoding is inherently designed to decode according to marginal likelihoods, which is exactly what JSA optimizes. Therefore MLS decoding helps JSA-SPG more significantly, while ``SPG separate training'' benefits less from MLS decoding.}

\RI{
JSA-SPG starts with fine-tuning the S2P model on 10 minutes of phoneme labels, which is then used to generate phoneme pseudo labels on the full training set.
In the second baseline, the G2P model (instead of the S2P model) is firstly fine-tuned on 10 minutes of phoneme labels and used to generate phoneme pseudo labels on the full training set, which are then be used to train S2P and P2G models.
The resulting S2P, P2G and G2P models are referred to as ``SPG training init from G2P''.
Further applying the JSA algorithm to jointly train the three models is found to bring no improvement and so we only report the result for applying ``SPG training init from G2P''.
Presumably, this is because the G2P fine-tuned in this way is overfit to the 10 minutes of phoneme labels, which is difficult to be further trained to be a good proposal.
``SPG training init from G2P'' shows worse results than JSA-SPG.
Note that in decoding, it is the hypothesized phonemes from S2P that are fed to P2G. But in  the training procedure in ``JSA-SPG training init from G2P'', the pseudo phoneme labels used to train P2G are from G2P. This causes some mismatch in training and decoding, and presumably explains the worse result of ``SPG training init from G2P'', compared to JSA-SPG.
}

\begin{table*}
    \caption{Performance comparison of different amounts of phoneme labels as supervised data in JSA-SPG training. MLS: marginal likelihood scoring.
    Except the column indicated by PER, other columns show WERs.}
    \label{tab:data volume}
    \begin{center}
    \begin{tabular}{l|cccc|cccc}
        \toprule
         \multirow{4}{*}{Amount of supervised data} & \multicolumn{4}{c|}{Polish} & \multicolumn{4}{c}{Indonesian} \\
         & \multirow{3}{*}{PER}  & \multicolumn{3}{c|}{WER} & \multirow{3}{*}{PER} & \multicolumn{3}{c}{WER}  \\
         \cline{3-5}  \cline{7-9}
         & & \multirow{2}{14mm}{decoding without LM} & \multirow{2}{11mm}{decoding with LM} & \multirow{2}{12mm}{decoding with \textit{MLS}} &   & \multirow{2}{14mm}{decoding without LM} & \multirow{2}{11mm}{decoding with LM} & \multirow{2}{12mm}{decoding with \textit{MLS}} \\
         & & & & & & & &  \\
         \midrule
         Unsupervised (i.e. zero-shot) & \multirow{2}{*}{50.24} & 13.43 & 6.20 & 5.05 & \multirow{2}{*}{32.71} & 11.09 & 3.74 & 2.80\\
        ~~~~ + P2G augmentation &  & 6.08 & 5.28    & 4.48 &  & 5.33 & 3.28 & 2.47 \\
        \midrule
        20 utterances (about 2 minutes) & \multirow{2}{*}{27.35} & 8.25 & 5.17 & 4.25 & \multirow{2}{*}{27.37} & 9.01 & 3.39 & 2.66\\
        ~~~~ + P2G augmentation &  & 5.31 & 4.78 & 3.96 &  & 5.45 & 3.04 & 2.47 \\
        \midrule
        100 utterances (about 10 minutes) & \multirow{2}{*}{\textbf{17.35}} & 8.19 & 4.65 & 3.93 & \multirow{2}{*}{\textbf{20.66}} & 9.04 & 3.26 & 2.47\\
        ~~~~ + P2G augmentation &  & \textbf{4.64} & \textbf{4.37} & \textbf{3.64} &  & \textbf{4.55} & \textbf{2.92} & \textbf{2.31} \\
        \bottomrule
    \end{tabular}
    \end{center}
\end{table*}

\textbf{Experiments with different amounts of supervised phoneme data.}
Table \ref{tab:data volume} shows ablation experiments with different amounts of supervised phoneme data. As the amount of supervised data increases, both PER and WER of the JSA-SPG model significantly decrease. Compared to 2 minutes of supervised data, with 10 minutes semi-supervised training, PER decreases by 36\% and WER by 8\% in Polish; Compared to unsupervised training, PER decreases by 65\% and WER by 18\%. There is also the same trend in the semi-supervised JSA-SPG experiments of the Indonesian. 

On the other hand, with reduced amounts of phoneme supervision such as only 2 minutes or even zero-shot, JSA-SPG obtains impressive results.
Notably, the zero-shot PERs of Polish and Indonesian by the Whistle-small model are 58\% and 46\% respectively. Without any phoneme supervision data, unsupervised JSA-SPG training leads to a significant reduction in PERs for both languages: a 13\% decrease for Polish and a 30\% decrease for Indonesian. The MLS decoding result of Indonesian (2.47\%) even surpasses that of subword fine-tuning (2.92\%) and approaches the result of phoneme fine-tuning (2.43\%). 
It should be emphasized that all 35 phonemes of the Indonesian are present in the Whistle phoneme set. We copy the corresponding weights for parameter initialization in the JSA-SPG training. For Polish, 31 phonemes appear in the Whistle phoneme set, while four phonemes do not. We randomly initialized the weights for those phonemes unseen in the Whistle phoneme set.  
This may account for the less-than-ideal performance of unsupervised Polish training. However, with the addition of 2 minutes of phoneme labels as supervision data, the WER from Polish JSA-SPG training (3.96\%) is lower than that of 130-hour full phoneme fine-tuning (4.30\%) and close to the result of subword fine-tuning (3.82\%). 

\RI{In the case that the target language contains phonemes not covered in the S2P backbone, we have two comments about the model’s behavior based on the results in Table \ref{tab:data volume}.
First, when phoneme-labeled training data that include these unseen phonemes are available—even in very limited amounts, such as only two minutes—the SPG model can still learn to recognize these unseen phonemes.
Second, in the extreme setting of zero-shot learning, where no phoneme-labeled data are provided at all, our experiments show that regions corresponding to unseen phonemes tend to be recognized as acoustically similar seen phonemes, e.g., recognizing /\texttoptiebar{\textrtaild\textrtailz}/ as /d \textrtailz/. Interestingly, such S2P errors can be partially corrected by the P2G component, since it is trained on noisy hypothesized phoneme sequences within the JSA-SPG framework. Such behavior may have important implications, which merit further exploration.
We conjecture that, in the context of constructing speech recognition systems, absolute precision in phonetic transcription may not be essential.}

\begin{table*}
    \caption{\RI{Resource Consumption of different decoding steps with the JSA-SPG model in the Polish experiment.
    RTF: real-time factor.}}
    \label{tab:resources}
    \begin{center}
    \begin{tabular}{c|l|ccc}
        \toprule
        \multirow{2}{*}{Step} & \multirow{2}{*}{Method} & CPU Memory & GPU Memory & \multirow{2}{*}{RTF} \\
         & & (GB) & (GB) &  \\
        \midrule
        1 & S2P inference & 0.58 & 1.00 & 0.008 \\
        2 & Beam search based P2G decoding & 1.74 & 1.04 & 0.004\\
        3 & WFST based P2G decoding & 0.30 & 0 & 0.003\\
        4 & MLS rescoring & 1.72 & 4.25 & 0.005 \\
        \bottomrule
    \end{tabular}
    \end{center}
\end{table*}

\RI{
\textbf{Experimental results on decoding complexity.}
Table \ref{tab:resources} summarizes experimental results on decoding complexity of the three decoding methods: ``decoding without LM'' (Step 1 and 2), ``decoding with LM'' (Step 1 and 3), and ``MLS decoding'' (Step 1, 3 and 4), respectively.
To ensure a fair comparison, all experiments were conducted with a batch size of 1 using a single GeForce RTX 3090 GPU.
Step 1, 2 and 4 were executed on the GPU, while Step 3, which employed WFST-based decoding, was run on the CPU with a single process. 
For memory measurements, we recorded the peak utilization values for both GPU and CPU, as memory usage fluctuates during decoding.
The main difference between ``decoding with LM'' and ``MLS decoding'' lies in the extra Step 4, i.e., calculating the MLS scores in the later method. 
The results in Table \ref{tab:resources} show that MLS decoding marginally increases inference complexity.
}

\section{Conclusion and future work}
\label{sec:conclusion}
In this paper, our aim is to achieve crosslingual speech recognition based on phonemes without pronunciation lexicons. By treating phonemes as discrete latent variables, modeling S2P and P2G together as a latent variable model (SPG), and introducing a G2P model as an auxiliary inference model, we utilize the JSA algorithm to jointly train these three networks. 
We refer to this new approach as JSA-SPG. Particularly, the S2P model is initialized from Whistle, a pre-trained phoneme-based multilingual S2P backbone.
This paper also proposes marginal likelihood scoring and P2G augmentation, which further \RI{enhance} the performance of JSA-SPG.
In crosslingual experiments with two languages (Polish and Indonesian), JSA-SPG method demonstrates remarkable performance. By utilizing merely 10 minutes of speech with phoneme labels,  it outperforms full data (130 and 20 hours) phonetic supervision. This effectively eliminates the necessity of using a PROLEX. 
Moreover, the JSA-SPG method surpasses crosslingual subword fine-tuning, and we take the language domain adaptation task as an example to show the bonus brought by JSA-SPG.
It is found that JSA-SPG significantly outperforms the standard practice of language model fusion via the auxiliary support of the G2P model. 

This paper presents some promising results along the JSA-SPG approach. 
There are interesting directions for future work.
First, the JSA-SPG method can be applied in phoneme-based pre-training to exploit a larger amount of data from more languages, even those languages without PROLEXs.
\PDF{Second, in our experiments, the S2P, P2G, and G2P models are all implemented—though not restricted to—using the CTC based models. Future work can explore alternative architectures as long as they are compatible with Algorithm \ref{alg:whistle-jsa}, such as LLM-based P2G models \cite{ma2025llm}.}
Third, JSA-SPG simultaneously produces S2P, P2G and G2P models, each of which potentially can be applied to a variety of speech processing tasks. Language domain adaptation is only one example. G2P is useful for text-to-speech. S2P may provide a new way for discrete tokenization for speech. 

\bibliographystyle{IEEEtran}
\bibliography{references.bib}

\vspace{-1cm}
\begin{IEEEbiography}[{\includegraphics[width=1in,height=1.25in,clip,keepaspectratio]{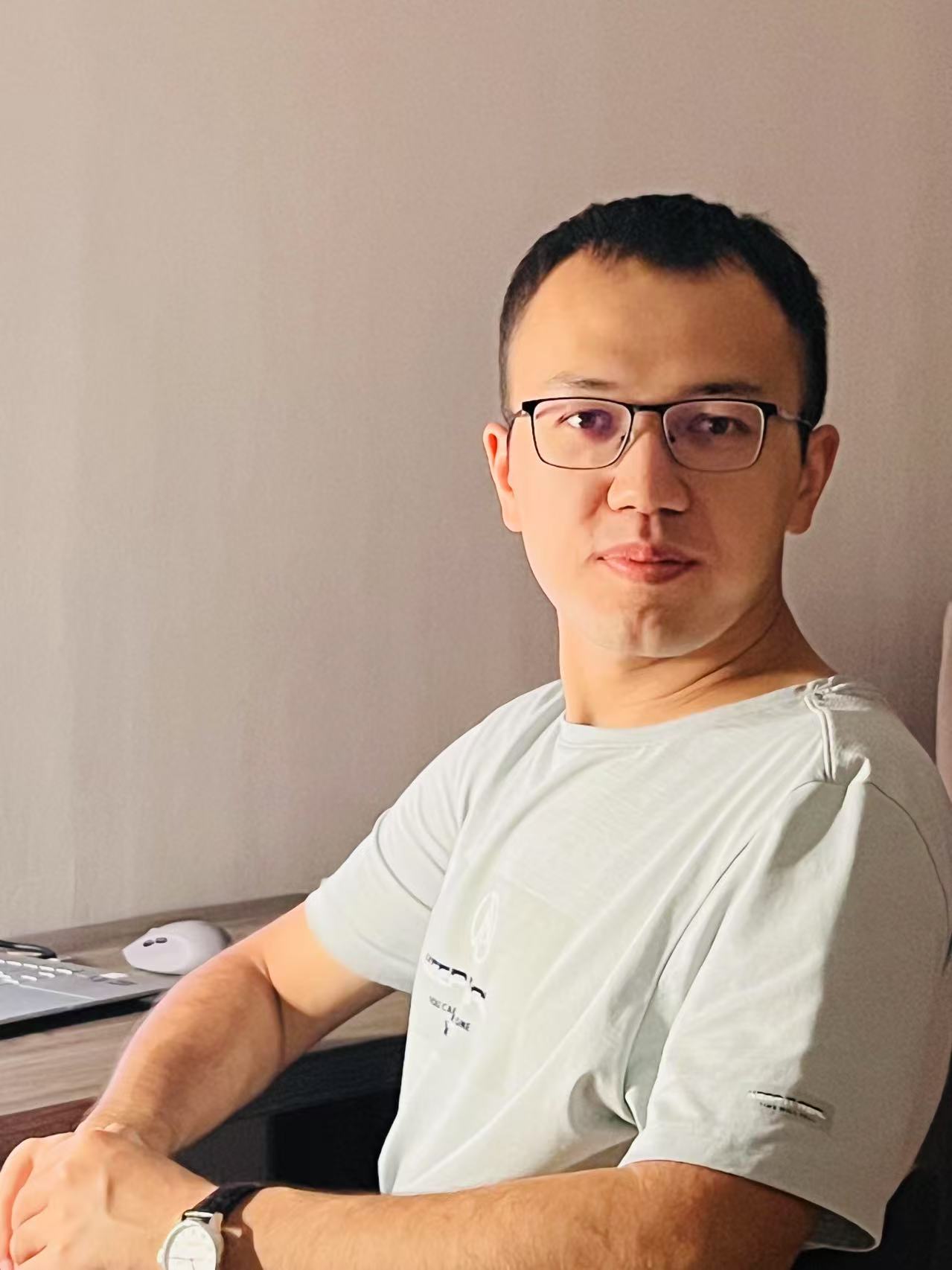}}]{Saierdaer Yusuyin}
received the B.E. degree in software engineering from Northwestern Polytechnical University, Xi'an,
China, in 2019. Since 2019, he has been working toward the Ph.D. degree with the School of Computer Science and Technology, Xinjiang University, under the supervision of Hao Huang and Zhijian Ou (during internship at THU-SPMI since January 2023). His research interests include multiliangual speech recognition and semi-supervised learning theory. 
\end{IEEEbiography}

\vspace{-1cm}

\begin{IEEEbiography}[{\includegraphics[width=1in,height=1.25in,clip,keepaspectratio]{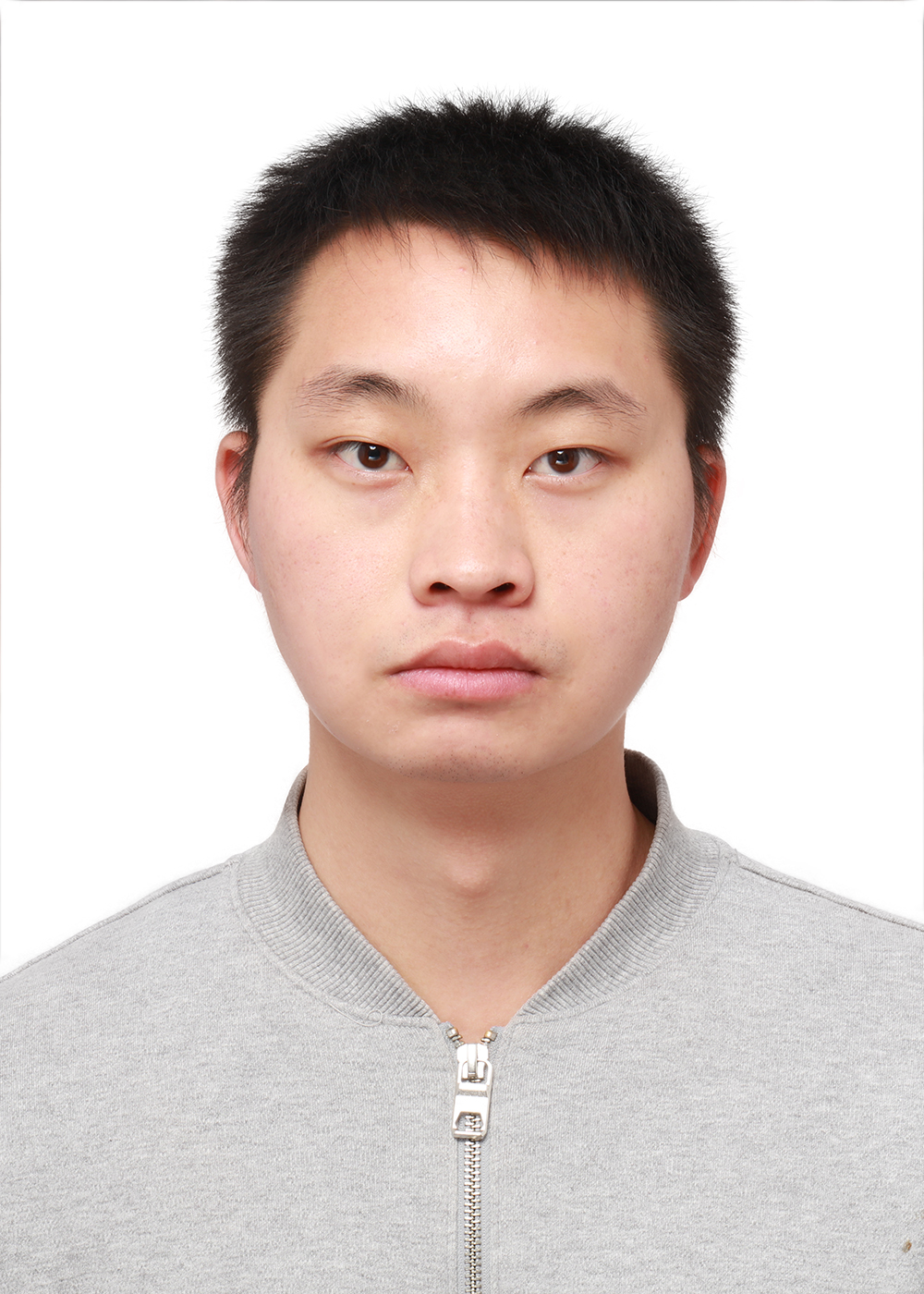}}]{Te Ma}
received the B.E. degree in electronic Information engineering from Chongqin University, Chongqin, China, in 2022. Since 2022, he has been working toward the master's degree with the School of Computer Science and Technology, Xinjiang University, under the supervision of Hao Huang  and Zhijian Ou (during internship at THU-SPMI since January 2023). His research interests include multilingual speech recognition and weakly supervised learning theory.
\end{IEEEbiography}

\newpage
\vspace{-1cm}

\begin{IEEEbiography}[{\includegraphics[width=1in,height=1.25in,clip,keepaspectratio]{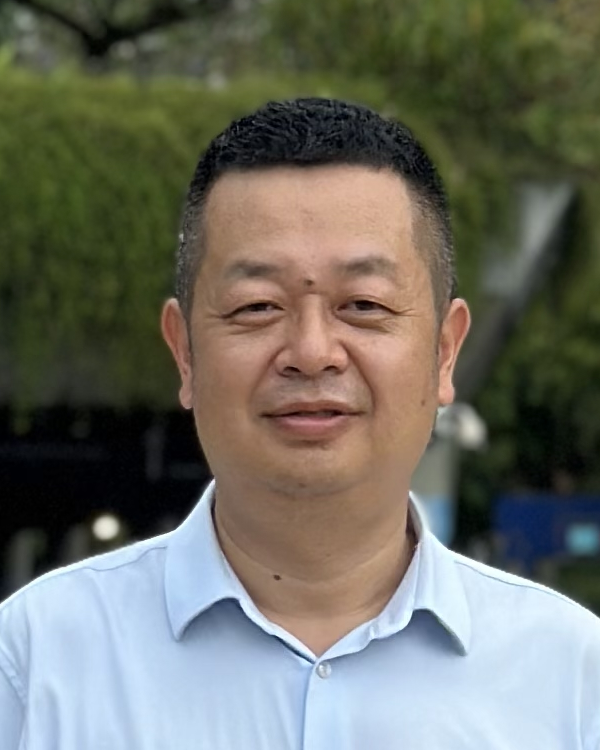}}]{Hao Huang (Member, IEEE)}
received the B.E. degree from Shanghai Jiao Tong University, Shanghai, China, in 1999, the M.E. degree from Xinjiang University, Urumqi, China, 2004, and the Ph.D. degree from Shanghai Jiao Tong University, in 2008, respectively. He is currently a Professor with the School of Computer Science and Technology, Xinjiang University. His current research interests include speech and language processing, and multi-media human-computer interaction. 
\end{IEEEbiography}

\vspace{-1cm}

\begin{IEEEbiography}[{\includegraphics[width=1in,height=1.25in,clip,keepaspectratio]{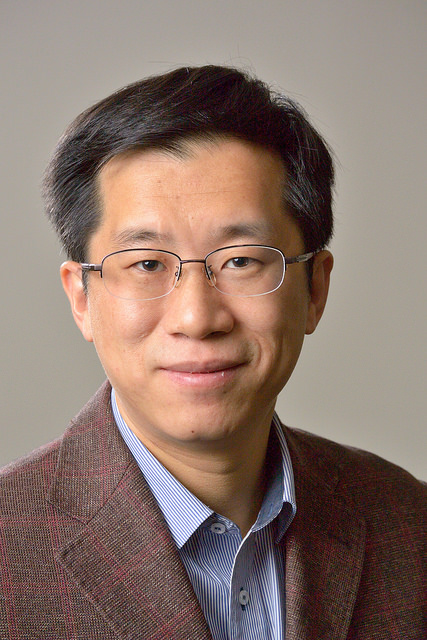}}]{Zhijian Ou (Senior Member, IEEE)}
received the Ph.D. degree from Tsinghua University, Beijing, China, in 2003. He is currently a Professor with the Department of Electronic Engineering, Tsinghua University, and Co-founder of TasiTech.  His research interests lie broadly in speech and language processing, with a particular emphasis on developing data-efficient approaches for speech recognition and dialog systems. He is an Senior Area Editor of IEEE/ACM TRANSACTIONS ON AUDIO, SPEECH AND LANGUAGE PROCESSING and SIGNAL PROCESSING LETTERS, an Editorial Board Member of COMPUTER SPEECH AND LANGUAGE, a Member of IEEE Speech and Language Processing Technical Committee, and was the General Chair of SLT 2021, EMNLP 2022 SereTOD Workshop, SLT 2024 FutureDial-RAG Challenge, Technical Program Chair of ISCSLP 2024, and Tutorial Chair of INTERSPEECH 2020. 
He has continuously led national research programs from Ministry of Science and Technology and Ministry of Education of China, National Natural Science Foundation of China (NSFC), and enterprise-funded projects with partners such as China Mobile, TasiTech, Meituan, Apple, Toshiba, IBM, Panasonic, and Intel. His work has been honored with four provincial and ministerial science and technology awards.
\end{IEEEbiography}


 




\vfill

\end{document}